\documentclass{appolb}
\usepackage{graphicx}
\usepackage{xspace}

\usepackage{graphicx}
\usepackage{subfig}
\usepackage{amssymb}
\usepackage{amsthm}
\usepackage{mathptmx}
\usepackage[utf8]{inputenc}
\usepackage[english]{babel}
\usepackage{nicefrac}
\usepackage{xcolor}
\usepackage[switch]{lineno}
\usepackage{siunitx}
\usepackage{chemformula}
\usepackage{float}
\DeclareSIUnit\barn{b}
\DeclareSIUnit\bar{bar}

\begin{document}

\title{Intermediate Mass Kaonic Atoms at DA$\Phi$NE}
\author{F Artibani
\address{Laboratori Nazionali di Frascati INFN Frascati, Italy}
\address{University of Roma Tre, Italy}
\\[3mm]
{F Clozza
\address{Laboratori Nazionali di Frascati INFN Frascati, Italy}
\address{University of Rome La Sapienza, Italy}
}
\\[3mm]
{M Bazzi, C Capoccia, A Clozza, L De Paolis, K Dulski, C Guaraldo, M Iliescu, A Khreptak, S Manti, F Napolitano, O Vazquez Doce, A Scordo, F Sgaramella, F Sirghi, A Spallone
\address{Laboratori Nazionali di Frascati INFN Frascati, Italy}
}
\\[3mm]
{M Cargnelli, J Marton, M T\"uchler, J Zmeskal 
\address{Stefan-Meyer-Institut f\"ur Subatomare Physik, Vienna, Austria}
}
\\[3mm]
{L Abbene, A Buttacavoli, F Principato
\address{Department of Physics and Chemistry (DiFC)—Emilio Segr\`e, University of Palermo, Palermo, Italy}
}
\\[3mm]
{D Bosnar, I Fri\v{s}\v{c}i\'c
\address{Department of Physics, Faculty of Science, University of Zagreb, Zagreb, Croatia}
}
\\[3mm]
{M Bragadireanu
\address{Horia Hulubei National Institute of Physics and Nuclear Engineering (IFIN-HH) M\u{a}gurele, Romania}
}
\\[3mm]
{G Borghi, M Carminati, G Deda, C Fiorini
\address{Politecnico di Milano, Dipartimento di Elettronica, Informazione e Bioingegneria, Milano, Italy}
}
\\[3mm]
{R Del Grande
\address{Excellence Cluster Universe, Technische Universi\"at M\"unchen Garching, Germany}
\address{Laboratori Nazionali di Frascati INFN Frascati, Italy}
}
\\[3mm]
{M Iwasaki
\address{RIKEN, Tokyo, Japan}
}
\\[3mm]
{P Moskal, S Nied\'{z}wiecki, M Silarski, M Skurzok
\address{Faculty of Physics, Astronomy, and Applied Computer Science, Jagiellonian University, Krak\'{o}w, Poland}
\address{Center for Theranostics, Jagiellonian University, Krakow, Poland}
}
\\[3mm]
{H Ohnishi, K Toho
\address{Research Center for Electron Photon Science (ELPH), Tohoku University, Sendai, Japan}
}
\\[3mm]
{D Sirghi, K Piscicchia
\address{Centro Ricerche Enrico Fermi - Museo Storico della Fisica e Centro Studi e Ricerche "Enrico Fermi", Roma, Italy}
\address{Laboratori Nazionali di Frascati INFN Frascati, Italy}
}
\\[3mm]
{C Curceanu
\address{Laboratori Nazionali di Frascati INFN Frascati, Italy}
}
}
\maketitle

\begin{abstract}
The SIDDHARTA-2 collaboration aims to measure for the first time the shift and width induced on the $1s$ level of kaonic deuterium by the strong interaction. In the preliminary phase to the experiment, a test run using a Helium-4 target was performed to optimize the performance of the full experimental apparatus. This preliminary study highlighted the possibility to measure transition lines coming from intermediate mass kaonic atoms, such as kaonic carbon and kaonic aluminum. In order to measure transitions where strong interaction is manifesting at higher energies, out of the energy range of the SIDDHARTA-2 apparatus, the collaboration is testing a new detector system which exploits a novel compound semiconductor, the Cadmium Zinc Telluride. Tests are now running at DA$\Phi$NE to study the performance of this detector, exploring the possibility to build a dedicated setup.
\end{abstract}
  
\section{Introduction}

Phenomenological models based on QCD \cite{cieply_pole_2016} and on various crucial observables, describe the behaviour of strong interactions in hadronic systems in the low energy limit. An important set of these observables can be derived by studying exotic atoms \cite{tomonaga_effect_1940}: systems in which a negatively charged particle (either a lepton or a hadron) is captured in an atomic orbit, replacing an atomic electron. The exotic particle is captured in a highly excited state and then de-excites towards lower lying states by emitting radiation \cite{Gotta:2004rq}. In this framework, kaonic atoms spectroscopy is a key experimental tool to probe the low-energy strong interaction with strangeness \cite{curceanu2019modern}. By measuring the shift and the width induced by the strong interaction on the lower levels, stringent constraints on the kaon-nucleon (K-N) and kaon-nuclei interactions can be set. With the recent development of precise and fast X-ray detectors, new experiments on kaonic atoms can be performed. The results of these studies give input to the phenomenological models describing the low-energy QCD with strangeness, in particular the K-N interactions as function of the nuclear density, with important contributions in particle and nuclear physics and astrophysics (hyperon puzzle \cite{tolos_strangeness_2020}).

The SIDDHARTA-2 experiment at DA$\Phi$NE \cite{Sirghi:2024sgq} is presently taking data with a challenging goal: to measure for the first time the shift and the width induced by the strong interaction on the $1s$ level of kaonic deuterium. A preliminary study using a helium-4 target aiming to optimize the performance of the apparatus \cite{sgaramella2024characterization}, showed how the de-excitation lines coming from some intermediate-mass kaonic atoms (such as kaonic carbon and kaonic oxigen) arising from the setup materials can be actually measured. The SIDDHARTA-2 collaboration is now testing new technologies with the aim to set up a new, dedicated apparatus to measure the strong interaction effects on these intermediate mass kaonic atoms.

\section{Experimental Setup}
\begin{figure}[H]
\centering
    \includegraphics[width=0.8\linewidth]{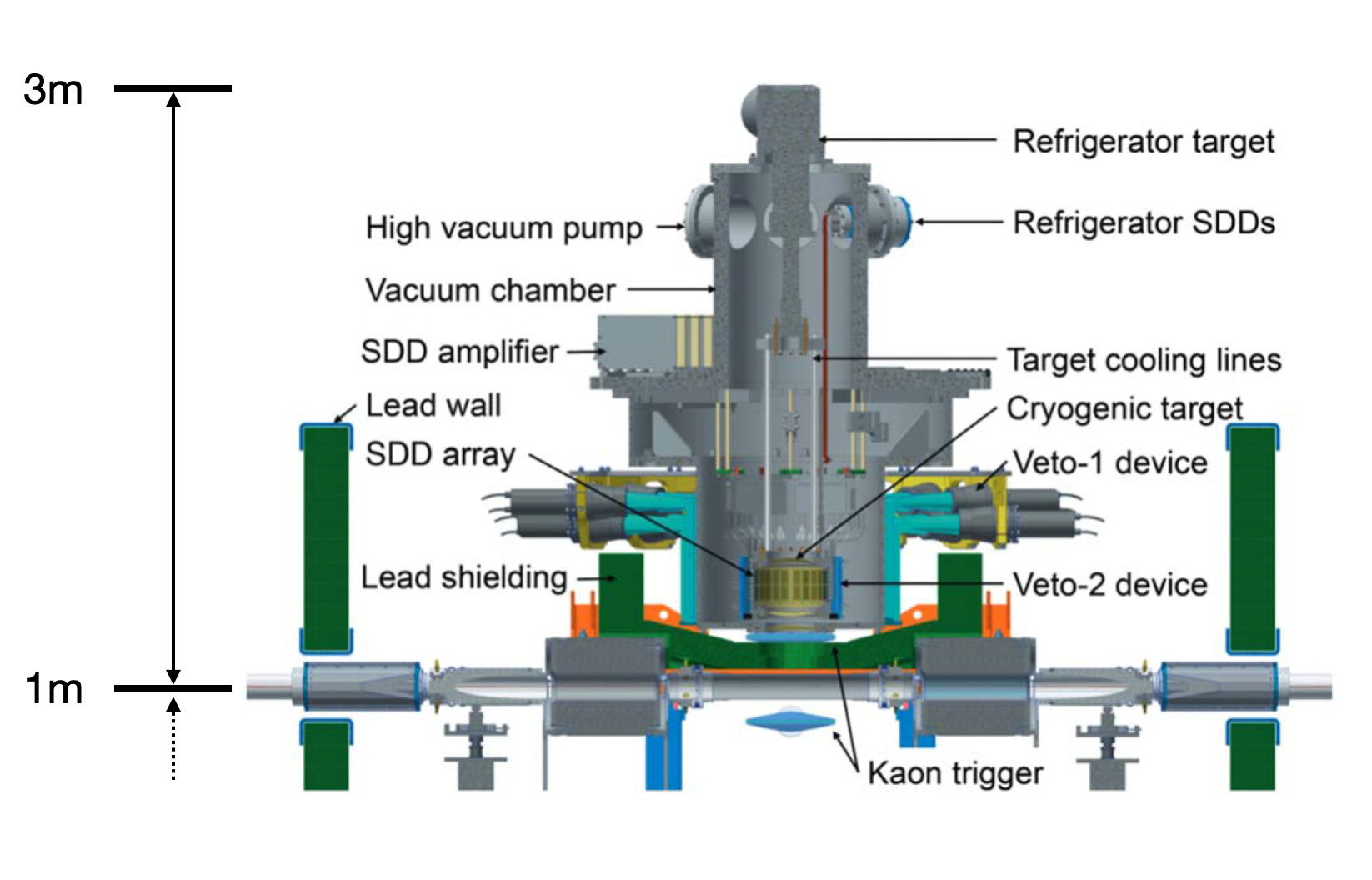}
  \caption{An overview of the experimental setup \cite{curceanu2019modern}. The whole system is installed at the $e^+$$e^-$ IP in DA$\Phi$NE.}
  \label{fig:siddsetup}
\end{figure}

The full SIDDHARTA-2 experimental setup was installed in the DA$\Phi$NE collider in Autumn 2022 \cite{Milardi:2018sih}. DA$\Phi$NE is an electron-positron collider working at the center of mass energy of the $\phi$ resonance (\SI{1.02}{\giga\electronvolt}), providing, via its decay, charged $K^+$/$K^-$ pairs with a branching ratio of $48.9\%$. The kaons are produced with a very low momentum (\SI{127}{\mega\electronvolt}/c), and can, therefore, be easily stopped inside a gaseous or solid target. The SIDDHARTA-2 experimental configuration is presented in Figure \ref{fig:siddsetup}, where the main elements are: the kaon trigger (KT), the mylar degrader, the luminosity monitor \cite{skurzok_characterization_2020}, the cylindrical vacuum chamber that contains the cryogenic target, the Silicon Drift Detector (SDDs) \cite{sgaramella_siddharta-2_2022, miliucci_large_2022, miliucci_silicon_2021, Khreptak_analysis_2023}, and three veto systems \cite{tuechler_charged_2018, tuchler_siddharta-2_2023}. The experimental setup was optimized with the help of a detailed Monte Carlo (MC) simulation using the GEANT4 CERN toolkit, to maximize the signal-to-background ratio during the data taking. More technical details on the experimental setup can be found in \cite{sirghi2023siddharta}.

\section{Kaonic Carbon and Aluminum lines}
\begin{figure}[H]
\centering
    \includegraphics[width=12.5cm]{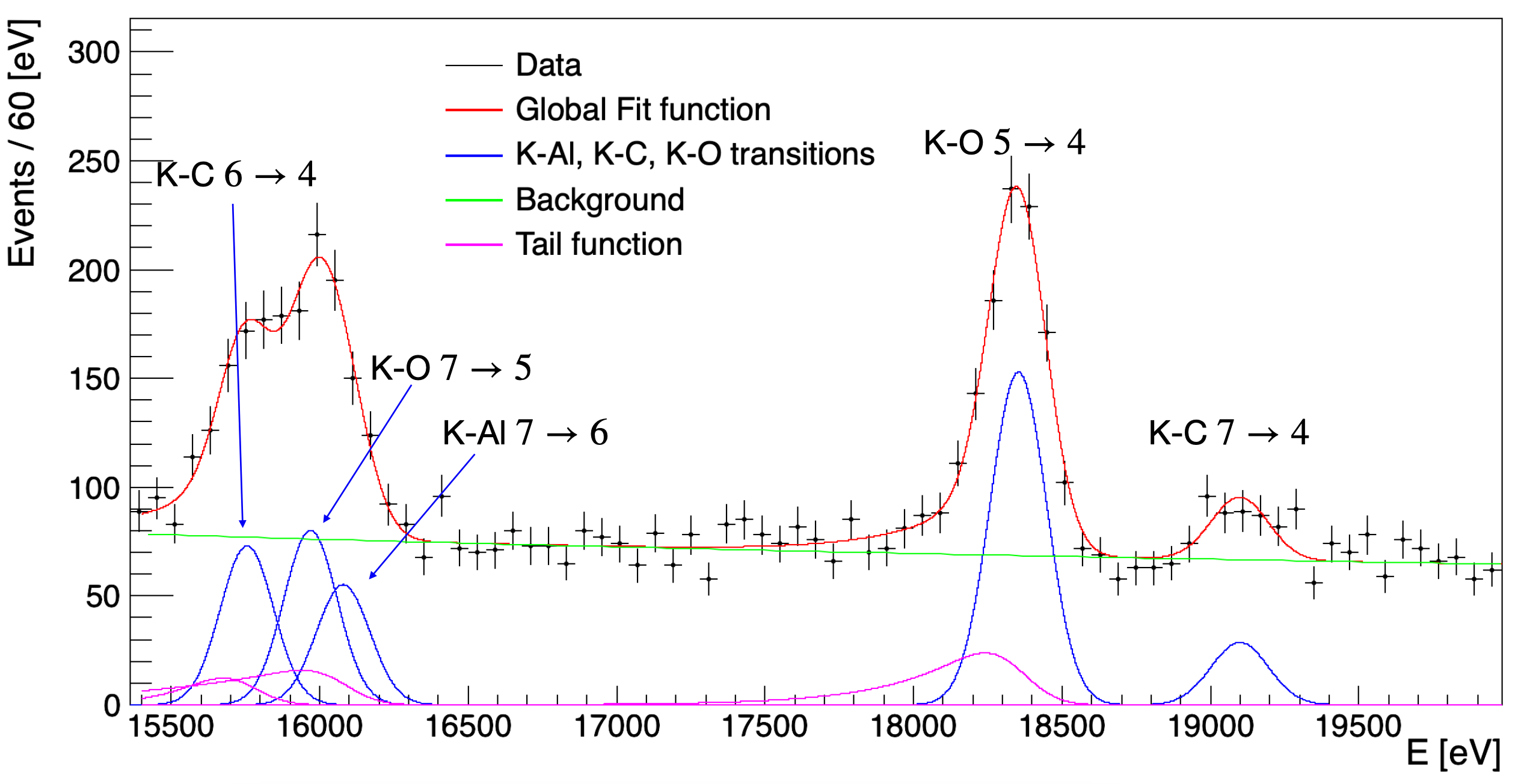}
  \caption{Fit to the kaonic helium-4 spectrum in the \SIrange[range-phrase=-, range-units=single]{15}{20}{\kilo\electronvolt} energy region.}
  \label{fig:spectrum}
\end{figure}

The recent measurement of kaonic helium-4 L-series transitions, which was carried out in 2023, confirmed the possibility of measuring other kaonic atoms transition lines at DA$\Phi$NE. The analysis of the measured spectrum was performed in the energy range \SIrange[range-phrase=-, range-units=single]{4}{12}{\kilo\electronvolt}, and in this region several kaonic carbon (K-C) and kaonic oxygen (K-O) lines were observed. These were the result of kaons interacting with the kapton walls (\ch{C_{22}H_{10}N_2O_5}) of the target cell of the experimental setup. Moving to higher energy in the spectrum, other K-C and K-O lines, together with a kaonic aluminum (K-Al) line, are visible and can be measured. With this purpose, a fit to the spectrum was performed. Each peak is described by a Gaussian function:
\begin{equation} \label{eq:gaussian}
    G(E) = \frac{Gain}{\sqrt{2\pi}\sigma} e^{\frac{-(E-E_0)^2}{2\sigma^2} } \, ,
\end{equation}
together with a tail component which describes the low energy contributions due to incomplete charge collection \cite{van2003implementation}:
\begin{equation} \label{eq:tailfunc}
    T(E) = \frac{Gain}{2\beta\sigma} e^{\frac{(E-E_0)}{\beta\sigma} + \frac{1}{2\beta^2} }\text{erfc}\left( \frac{E-E_0}{\sqrt{2}\sigma} + \frac{1}{\sqrt{2}\beta} \right) \, ,
\end{equation}
where $\beta$ is the slope parameter of the tail function. The result of the fit is shown in Figure \ref{fig:spectrum}. The measured values of the energies of the transitions, together with their QED calculated ones, are reported in Table \ref{tab:kaonicvalues}. The electromagnetic values have been calculated with vacuum polarization and recoil corrections referring to \cite{gotta1999balmer, karshenboim2006uehling, karshenboim2006vacuum}, and the results were then compared with the values previously calculated for the KEK-KpX \cite{iwasaki1997observation}, KEK-E570 \cite{okada2007precision} and DEAR \cite{zmeskal2005dear} experiments.

\begin{table} [H]
    \centering
    \begin{tabular}{ccc}
        Transition & E$^{\text{QED}}$ [eV]& E$^{\text{meas}}$ [eV]\\
        \hline
        K-C $6\rightarrow4$ & 15759.4 & 15756 $\pm$ 23 (stat) $\pm$ 9 (syst)\\
        K-O $7\rightarrow5$& 15973.3 & 15970 $\pm$ 23 (stat) $\pm$ 9 (syst)\\
        K-Al $7\rightarrow6$& 16088.3 & 16080 $\pm$ 16 (stat) $\pm$ 9 (syst)\\
        K-O $5\rightarrow4$ & 18370.5 & 18353 $\pm$ 11 (stat) $\pm$ 9 (syst)\\
        K-C $7\rightarrow4$ & 19101.0 & 19099 $\pm$ 32 (stat) $\pm$ 9 (syst)\\
    \end{tabular}
    \caption{Kaonic carbon, oxygen and aluminum transition lines with their respective nominal and measured energies.}
    \label{tab:kaonicvalues}
\end{table}
The systematic uncertainties were estimated by measuring the energy residual of the Bismuth L$_{\gamma1}$ line in the raw spectrum, which has a nominal value of \SI{15248}{\eV} \cite{kortright2001x} against a measured value of (15239.10 $\pm$ 0.68) eV. Therefore, the measured values are in good agreement with the theoretical ones inside the error bars. Nonetheless, the SDDs used by the SIDDHARTA-2 experiment are not optimally designed and calibrated in the \SIrange[range-phrase=-, range-units=single]{15}{20}{\kilo\electronvolt} energy range, since the whole experimental setup was conceived to work at its best in the kaonic deuterium region of interest (\SIrange[range-phrase=-, range-units=single]{4}{12}{\kilo\electronvolt}). This measurement did not highlight any significant shift of these lines induced by the strong interaction. This should appear, according to \cite{friedman_density-dependent_1994}, for higher energy transitions of these kaonic atoms. This study emphasized that, in order to perform such measurements, a dedicated setup with a new detector designed to work in the \SIrange[range-phrase=-, range-units=single]{40}{400}{\kilo\electronvolt} energy range is needed.

\section{CZT detector for Intermediate Mass kaonic Atoms}
\begin{figure}[htb]
\centerline{
\includegraphics[width=12.5cm, height=0.6\textwidth]{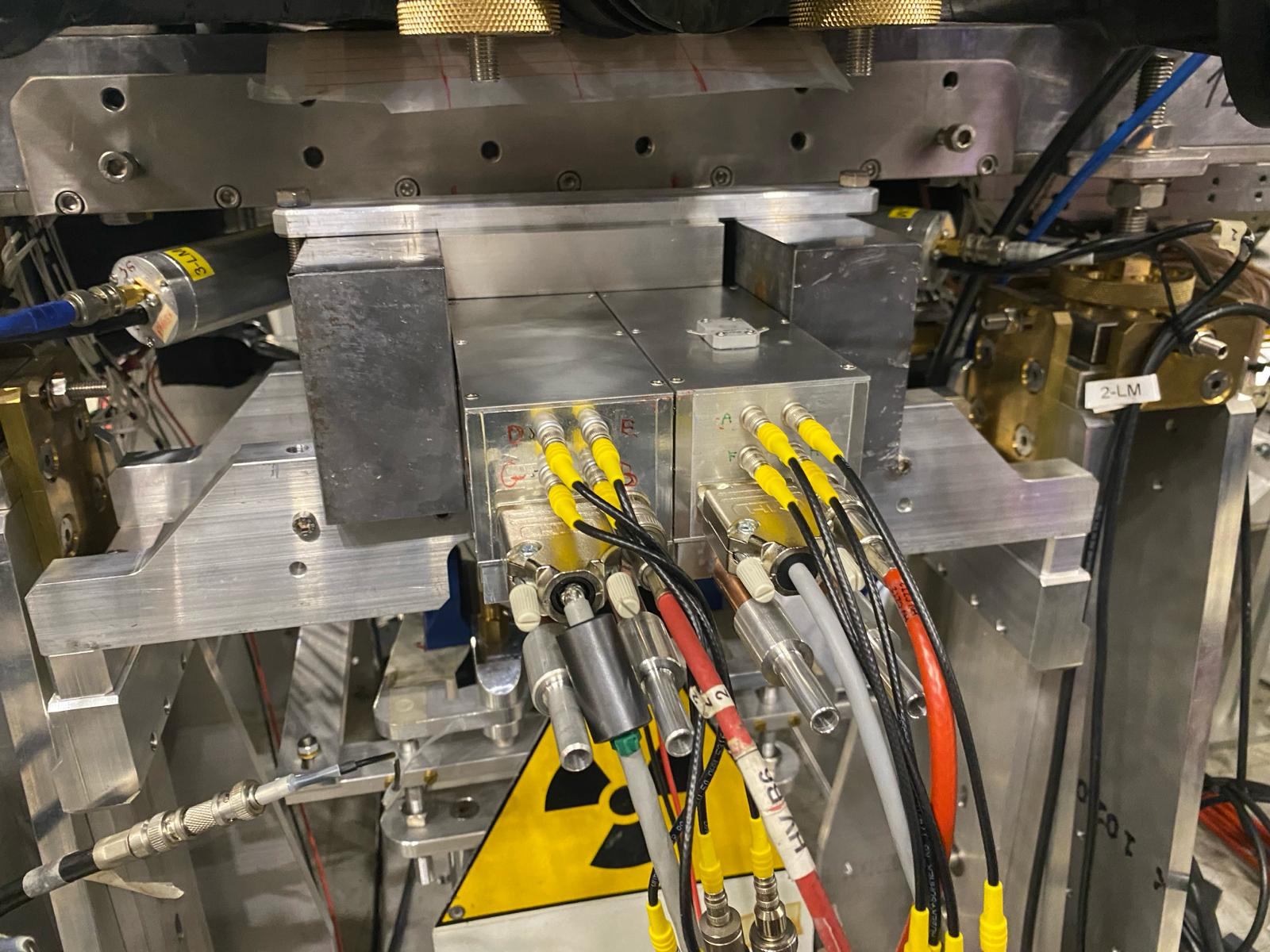}}
\caption{Picture of the CZT detectors installed in DA$\Phi$NE.}
\label{fig:czt}
\end{figure}

In recent times, a considerable improvement in radiation detection technologies was achieved. In the framework of X-ray and $\gamma$-ray detectors, compound semiconductors are the most promising, given the possibility to grow crystals with dedicated physical properties for any application. In particular, cadmium-zinc-telluride (CdZnTe, CZT) based detectors are strong candidates for experiments aiming to measure transitions of intermediate mass kaonic atoms, such as K-Al and K-C. In fact, in the energy region of interest for these transitions (tens to hundreds keV), CZT detectors exhibit good efficiency, room temperature energy resolution and a fast timing, thus making this novel material excellent to produce new versatile and compact detection systems. Such detectors were already used in the fields of medical imaging \cite{iniewski_czt_2014} and astrophysics \cite{tang_cadmium_2021}. A prototype of CdZnTe based detector was tested for the first time in a collider by the SIDDHARTA-2 collaboration \cite{abbene_potentialities_2023, abbene_new_2023, scordo_cdznte_2024}, after accurate studies on the high-rate performances \cite{abbene_high-rate_2015}. The first tests in DA$\Phi$NE studied the feasibility to perform intermediate mass kaonic atoms, especially K-Al and K-C, obtaining promising results reported in \cite{scordo_cdznte_2024}. A new upgraded CZT detector, built by the SIDDHARTA-2 collaboration, features two modules with four custom $13 \times 15 \times 5$ mm$^3$ CdZnTe quasi-hemispherical detectors provided by REDLEN Technologies, connected to analog charge-sensitive preamplifiers (CSPs) and digital pulse processing (DPP) readout electronics, accurately described in \cite{abbene_development_2017, abbene_room-temperature_2020}. The detectors and the preamplifiers are placed inside a \SI{5}{\milli\meter} thick aluminum box, with an entrance window of \SI{0.2}{\milli\meter} shown in Figure \ref{fig:czt}.

The apparatus is now installed in the longitudinal plane of the DA$\Phi$NE interaction point, in such a way to not interfere with the SIDDHARTA-2 experimental setup and is taking data. The main goal of this run is a preliminary test for the measurement of K-Al and K-C shifts and widths on the $n=3$ and $n=2$ levels, respectively, while exploring the possibility of performing new dedicated experiment to measure intermediate mass kaonic atoms at DA$\Phi$NE or at J-PARC in the future.

\section{Conclusions}
In this paper the measurement of kaonic atoms transition lines appearing in the high energy region (\SIrange[range-phrase=-, range-units=single]{15}{20}{\kilo\electronvolt}) of the spectrum acquired by the SIDDHARTA-2 experiment during the Helium-4 run is reported. The analysis of these data delivered new measurements, reported in Table \ref{tab:kaonicvalues}, of the kaonic carbon, kaonic aluminum and kaonic oxygen purely electromagnetic transition lines. This study highlighted the feasibility of measuring transitions where strong interaction is present of intermediate mass kaonic atoms at DA$\Phi$NE. While exploring the possibility to perform new kaonic atoms experiments after the end of the kaonic deuterium SIDDHARTA-2 run \cite{curceanu_frontiers_2023}, the collaboration is now testing a new CdZnTe based detector which is designed to perform such measurements. The results of these future experiments can provide theoreticians with new valuable input data for phenomenological models of the low energy QCD with strangeness.

\section*{Acknowledgments}
We thank H. Schneider, L. Stohwasser, and D. Pristauz-Telsnigg from Stefan Meyer-Institut for their fundamental contribution in designing and building the SIDDHARTA-2 setup. We thank as well the INFN, INFN-LNF and the DA$\Phi$NE staff in particular to Dr. Catia Milardi for the excellent working conditions and permanent support.
Catalina Curceanu acknowledge University of Adelaide, where part of this work was done (under the George Southgate fellowship, 2023).

Part of this work was supported by the Austrian Science Fund (FWF): [P24756-N20 and P33037-N]; the EXOTICA project of the Minstero degli Affari Esteri e della Cooperazione Internazionale, PO22MO03; the Croatian Science Foundation under the project IP-2018–01-8570; the EU STRONG-2020 project (Grant Agreement No. 824093); the EU Horizon 2020 project under the MSCA (Grant Agreement 754496); the Japan Society for the Promotion of Science JSPS KAKENHI Grant No. JP18H05402; the SciMat and qLife Priority Research Areas budget under the program Excellence Initiative - Research University at the Jagiellonian University, and the Polish National Agency for Academic Exchange (Grant No. PPN/BIT/2021/1/00037); the EU Horizon 2020 research and innovation programme under project OPSVIO (Grant Agreement No. 101038099).

\bibliography{ref}

\end{document}